\documentclass[epj]{svjour}

\usepackage{graphicx}
\usepackage{fancyhdr}
\def\makeheadbox{{
 }}
\def\mytitle{My title} 
\def\myauthors{My name}  
\def\mytype{My type of session}
\def\mysession{My session}

\def\mytitle{Radiative corrections to neutralino annihilation in the A-funnel}
\def\myauthors{Bj\"orn Herrmann}
\def\mytype{Contributed Talk}    
\def\mysession{Cosmology}

\begin{document}

\title{Effect of SUSY-QCD corrections to neutralino annihilation on the cold dark matter relic density in the Higgs funnel}

\author{Bj\"orn Herrmann 
}                    

\institute{Laboratoire de Physique Subatomique et de Cosmologie\\
	   Universit\'e Joseph Fourier / CNRS-IN2P3 / INPG, 
	   53 Avenue des Martyrs, 38026 Grenoble, France\\
}

\date{September 5, 2007}

\abstract{
We present a complete calculation of the QCD and SUSY-QCD corrections to
neutralino pair annihilation into bottom quark-antiquark pairs through exchange
of a pseudoscalar Higgs boson, which is the dominant process in the cosmological
{\it A-funnel} region of the mSUGRA model. We present numerical predictions for 
the annihilation cross section and discuss the influence of the correction terms 
on the cold dark matter relic density with respect to recent cosmological data.
} 

\maketitle

\section{Motivation}
\label{sec1}

\vspace*{-12.3cm}LPSC 07-100\vspace*{12.3cm}

Thanks to the recent WMAP mission and further cosmological observations, the
matter and energy decomposition of our Universe is known today with
unprecedented precision \cite{ref1}. Direct evidence for the existence of Cold
Dark Matter (CDM) is accumulating \cite{ref2}, and its relic density $\Omega_{\rm
CDM}$ can be constrained to rather narrow range \cite{ref4}
\begin{equation}
	0.094 \le \Omega_{\rm CDM}h^2 \le 0.136
\label{eq1}
\end{equation}
at 95\% confidence level. $h$ denotes the present Hubble
expansion rate $H_0$ in units of 100 ${\rm km}~{\rm s}^{-1}{\rm Mpc}^{-1}$.

Although the nature of dark matter remains still unknown, extensions of the
Standard Model (SM) of particle physics provide interesting candidates for these
Weakly Interacting Massive Particles (WIMPs). In Supersymmetry (SUSY), a natural
candidate is the Lightest Supersymmetric Particle (LSP), which is stable, if
$R$-parity is conserved. It is usually the lightest of the four neutralinos,
denoted $\tilde{\chi}_1^0$ or shortly $\chi$.

The Minimal Supersymmetric SM (MSSM) is governed by 124 a priori free
soft SUSY-breaking parameters, which are often restricted to five universal
parameters that are imposed at the unification scale and can be constrained
using data from high-energy colliders. As the neutralino relic density also 
depends on these parameters, its computation is another powerful tool to put
constraints on the parameter space and provide complementary information, in
particular at high energies and masses that would otherwise not be accessible at
colliders.

To evaluate the number density $n$ of the relic particle, one has to solve the
Boltzmann equation
\begin{equation}
	\frac{dn}{dt} = -3Hn - \langle\sigma_{\rm eff}v\rangle 
		\left( n^2 - n_{eq}^2 \right) 
\label{eq2}
\end{equation}
with the Hubble rate $H$ and the thermal equilibrium density $n_{eq}$. The
number density $n$ is directly related to the relic density $\Omega_{\rm
CDM}h^2 = m_{\chi} n / \rho_c \propto \langle\sigma_{\rm eff}v\rangle^{-1}$,
where $m_{\chi}$ is the LSP mass, $\rho_c=3 H_0^2 / (8\pi G_N)$ is the critical
density of our Universe, and $G_N$ is the gravitational constant \cite{ref5}.
The effective cross section $\sigma_{\rm eff}$ involves all (co)annihilation
processes of the relic particle $\chi$ into SM particles, and
$\langle\sigma_{\rm eff}v\rangle$ signifies the thermal 
average of its non-relativistic expansion ($v\ll c$).

Several public codes perform a calculation of the dark matter relic density
within supersymmetric models. The most developed and most popular ones are {\tt
DarkSUSY} \cite{ref7} and {\tt micrOMEGAs} \cite{ref8}. All relevant
(co)anni\-hilation processes are implemented in these codes, but for most of them
no (or at least not the complete) higher order corrections are included. However,
due to the large magnitude of the strong coupling constant, QCD and SUSY-QCD
corrections are bound to affect the annihilation cross section in a significant
way. They may even be enhanced logarithmically by kinematics or in certain
regions of the parameter space.

It is the aim of this work to study these corrections for the neutralino-pair
annihilation into a bottom quark-antiquark pair through exchange of a
pseudoscalar Higgs boson $A^0$. This process dominates in the so-called {\it
A-funnel} region of the mSUGRA parameter space at large $\tan\beta$, which is
theoretically favoured by the unification of Yukawa couplings in Grand
Unification Theories (GUTs) \cite{ref9}. Supposing a WIMP mass of 50 - 70 GeV,
this process has been claimed to be compatible with the gamma-ray excess
observed by the EGRET satellite in all sky directions \cite{ref10}. However, the
corresponding scenarios may lead to antiproton overproduction, so that they
would not be compatible with the observed antiproton flux \cite{ref11}.

\section{Analytical results}
\label{sec2}

As we focus here on the phenomenological aspects of this work, the
analytical calculation is only briefly sketched. For a more detailed discussion
we refer the reader to Ref. \cite{ref0}.
\begin{figure}
	\includegraphics[scale=0.34]{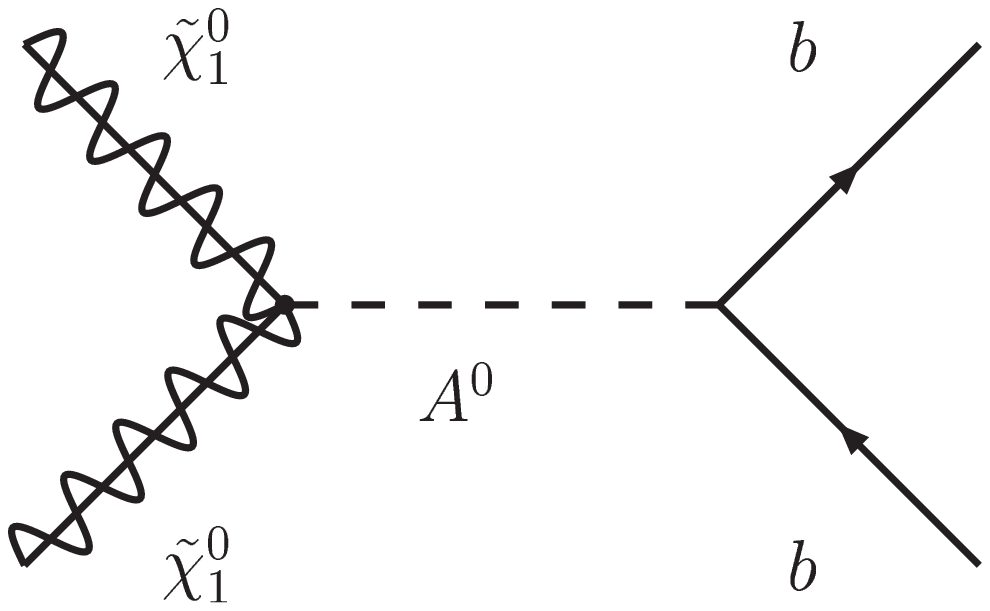}
	\hfill
	\includegraphics[scale=0.34]{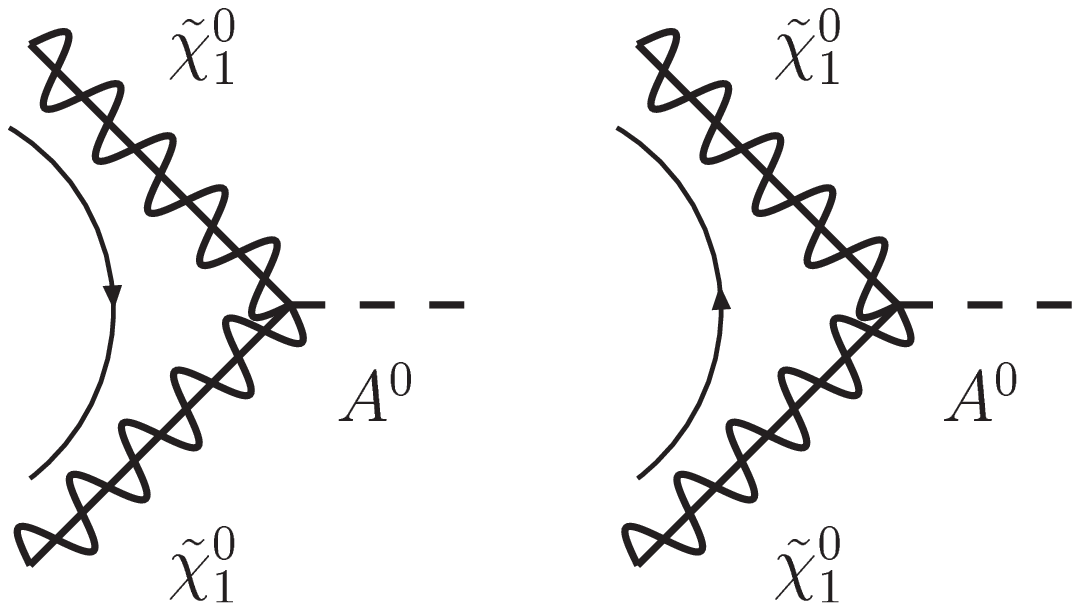}
\caption{The tree level diagram for the process $\chi\chi\to A^0\to b\bar{b}$,
	 indicating the two possible fermionic orientations of the Majorana 
	 initial state.}
\label{fig1}
\end{figure}

\begin{figure}
	\includegraphics[scale=0.4]{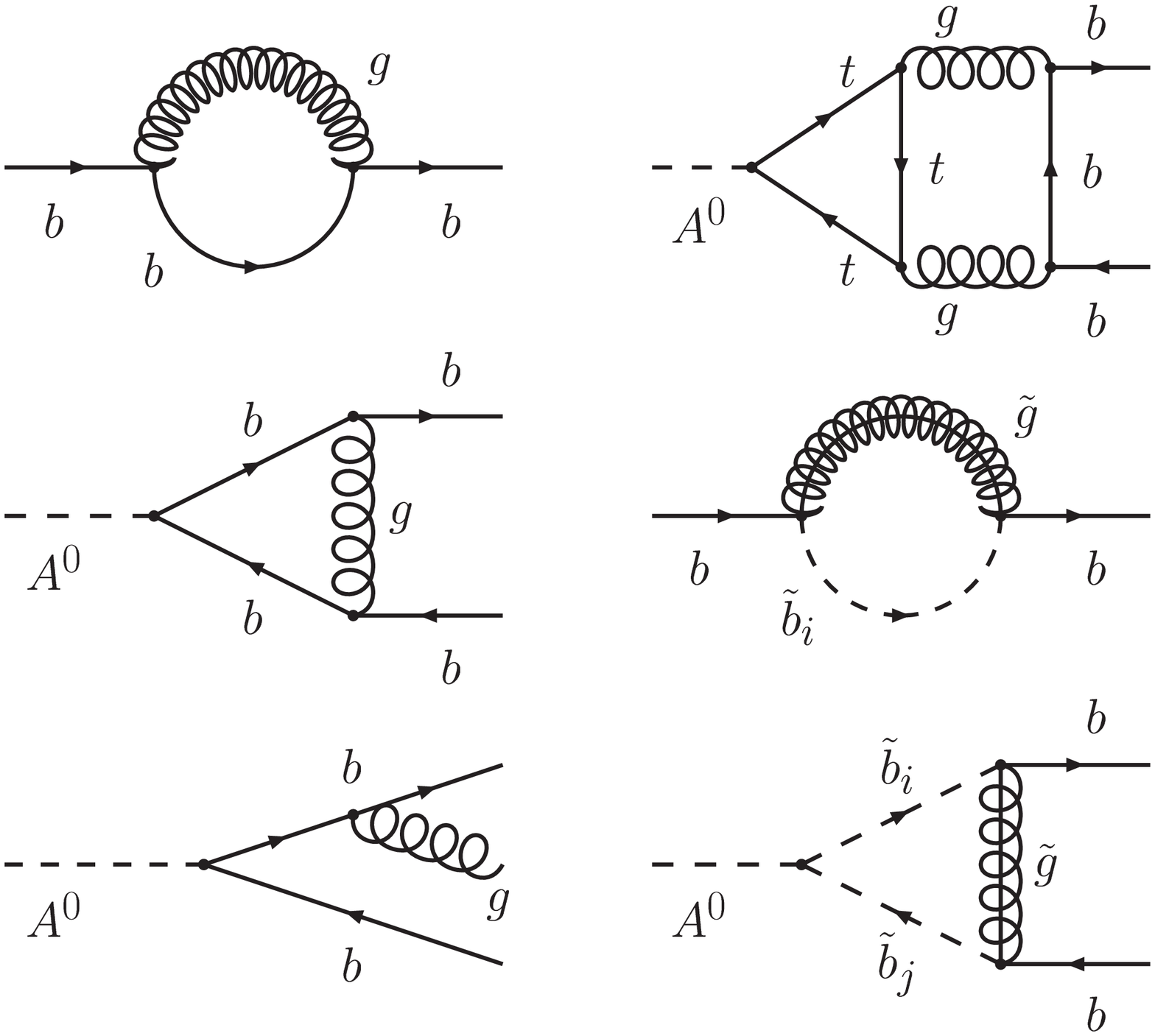}
\caption{QCD (left), top-quark loop (right top), and SUSY-QCD (right centre and
	 bottom) correction diagrams for the annihilation process $\chi\chi\to
	 A^0\to b\bar{b}$.} 
\label{fig2}
\end{figure}

In the calculation of the leading order (LO) amplitude, we have to make sure to
anti-symmetrize the neutralino pair in the initial state properly by taking into
account both possible fermionic orientations shown on the right-hand side of
Fig. \ref{fig2}, leading to a factor $\sqrt{2}$ in the amplitude with respect
to the amplitude of one fixed orientation. Denoting by $\sqrt{s}$ the
centre-of-momentum energy and $\beta_{\chi}=v/2$ and $\beta_b$
the neutralino and $b$-quark velocities, the leading order cross section can
then be written as 
\begin{equation}
	\sigma_{\rm LO} v = \frac{1}{2} \frac{\beta_b}{8\pi s}
	\frac{N_C g^2 T_{A11}^2 h_{Abb}^2 s^2}{|s-m_A^2+im_A\Gamma_A|^2}.
\label{eq3}		
\end{equation}
It is proportional to the inverse of the flux factor $sv$, the integrated
two-particle phase space $s\beta_b/(8\pi s)$, the number of quark colours
$N_C=3$ and the squares of the weak coupling constant $g$, a neutralino mixing
factor 
\begin{eqnarray}
	T_{Aij} &=& \frac{1}{2}\big( N_{2j}-\tan\theta_W N_{1j} \big)
	\big(N_{4i}\cos\beta-N_{3i}\sin\beta \big) \nonumber \\
	& & + (i\leftrightarrow j) ,
\label{eq4}
\end{eqnarray} 
the bottom quark mass through the Yukawa coupling $h_{Abb}=-g m_b \tan\beta /
(2m_W)$, and the Higgs boson propagator. The $N_{ij}$ in Eq. (\ref{eq4}) are the
entries of the matrix $N$ that diagonalizes the neutralino mass matrix. The
non-relativistic expansion of Eq. (\ref{eq4}) is performed by expanding the
squared centre-of-momentum energy $s$ in powers of $v^2$, 
\begin{equation}
	s \doteq 4m_{\chi}^2 \left(1 + \frac{v^2}{4} \right) + {\cal O}(v^4)
\end{equation}
and is in agreement with the results given in Ref. \cite{ref6}.

The next-to-leading order (NLO) annihilation cross section can be written as 
\begin{equation}
	\sigma_{\rm NLO} = \sigma_{\rm LO} \big( 1 + \Delta_{\rm QCD} +
		\Delta_{\rm top} + \Delta_{\rm SUSY} \big) ,
\label{eq5}	
\end{equation}
where $\Delta_{\rm QCD}$, $\Delta_{\rm top}$ and $\Delta_{\rm SUSY}$ 
relate to the QCD, top-quark loop, and SUSY-QCD correction diagrams shown in
Fig. \ref{fig2}. The same relation also holds for the Higgs boson width
$\Gamma_A$, for which the same correction diagrams are involved. Using 
standard methods, we compute the QCD correction $\Delta_{\rm QCD}$ at ${\cal
O}(\alpha_s)$, coming from the gluon exchange and real emission diagrams, which
agrees with the known result for pseudoscalar Higgs boson decays in the on-shell
scheme \cite{ref12}. In the limit $m_b^2 \ll s$, this correction develops a 
logarithmic mass singularity,
which can be resummed to all orders using the renormalization group, i.e.
replacing $m_b$ with the running mass $\bar{m}_b(s)$ in the Yukawa coupling
$h_{Abb}$ \cite{ref13}. The remaining finite QCD corrections in the $\overline{\rm MS}$
scheme are known up to ${\cal O}(\alpha_s^3)$,
\begin{eqnarray}
	\Delta_{\rm QCD} &=& \frac{\alpha_s(s) C_F}{\pi} \frac{17}{4} 
		+ \frac{\alpha_s(s)^2}{\pi^2} \big( 35.94 - 1.36 n_f \big) \nonumber\\
	& & + \frac{\alpha_s(s)^3}{\pi^3} \big( 164.14 - 25.77 n_f + 0.259 n_f^2
	\big), \quad
\end{eqnarray}
where $n_f$ denotes the number of flavours \cite{ref14}.

A separately gauge-independent correction at order ${\cal O}(\alpha_s^2)$ is
induced by the top-quark loop diagram shown in Fig. \ref{fig2}. Its contribution
\cite{ref15}
\begin{equation}
	\hspace*{-2mm}
	\Delta_{\rm top} = \frac{1}{\tan^2\beta} \frac{\alpha^2_s(s)}{\pi^2} 
		\left[ \frac{23}{6} - \log\frac{s}{m_t^2} 
		+ \frac{1}{6} \log^2\frac{\bar{m}_b^2}{s} \right]		
\end{equation}
can be important at small values of $\tan\beta$. However, it is largely
suppressed in the Higgs funnel region considered here, as shown in the centre 
panel of Fig. \ref{fig3}.

In SUSY, the sbottom-gluino exchange diagrams shown in Fig. \ref{fig2} give rise 
to additional corrections at ${\cal O}(\alpha_s)$. The self-energy diagram leads 
to the mass renormalization \cite{ref9}
\begin{equation}
	\Delta m_b = \frac{\alpha_s(s) C_F}{\pi} \frac{m_{\tilde{g}}}{2}
		\big( A_b - \mu\tan\beta \big) 
		I(m^2_{\tilde{b}_1},m^2_{\tilde{b}_2},m^2_{\tilde{g}})
\label{eq11}
\end{equation}
where $I(m_1,m_2,m_3) = C_0(0,0,0;m_1,m_2,m_3)$ is the 3-point function at zero
external momentum. In this low-energy (LE) limit, and neglecting $A_b$ with
respect to the $\tan\beta$-enhanced $\mu$, the vertex correction equals the mass
renormalization \cite{ref16} up to a factor $1/\tan^2\beta$, so that the total
SUSY correction becomes
\begin{eqnarray}
	\Delta_{\rm SUSY}^{\rm (LE)} &=& \frac{\alpha_s(s) C_F}{\pi}
		\frac{1+\tan^2\beta}{\tan^2\beta} m_{\tilde{g}} 
		\mu \tan\beta \nonumber \\
	& & \times\ I(m_{\tilde{b}_1},m_{\tilde{b}_2},m_{\tilde{g}}) .
\end{eqnarray}
As $A_b$ may be of similar size as $\mu\tan\beta$, its contribution must also be
resummed. This is effectively done by replacing \cite{ref17}
\begin{equation}
	\lim_{A_b\to 0}\Delta m_b~\rightarrow~
	\frac{\lim_{A_b\to 0}\Delta m_b}{1+\lim_{\mu\tan\beta\to 0}\Delta m_b} .
\end{equation}
Finally, our result agrees with the those in Refs.
\cite{ref18}, and we implement the finite ${\cal O}(m_b,s,1/\tan^2\beta)$
remainder as described in Ref. \cite{ref17}.

\section{Numerical discussion}
\label{sec3}

For our numerical study of the impact of the QCD, top-quark loop, and SUSY-QCD
corrections discussed above, we place ourselves in a minimal supergravity
(mSUGRA) scenario with five parameters $m_0$, $m_{1/2}$, $A_0=0$, $\tan\beta$, 
and sgn($\mu$) at the grand unification (GUT) scale. The weak-scale MSSM 
parameters are then determined through renormalization group running using 
{\tt SPheno} \cite{ref20}, and the physical Higgs and SUSY masses with 
{\tt FeynHiggs} \cite{ref21}. For the SM inputs, i.e. masses and widths of the
and electroweak gauge bosons and quarks, the coupling constants, the entries of
the CKM-matrix, and the $CP$-violating phase, we refer the reader to Ref. \cite{ref22}.
\begin{figure*}
	\includegraphics[scale=0.27]{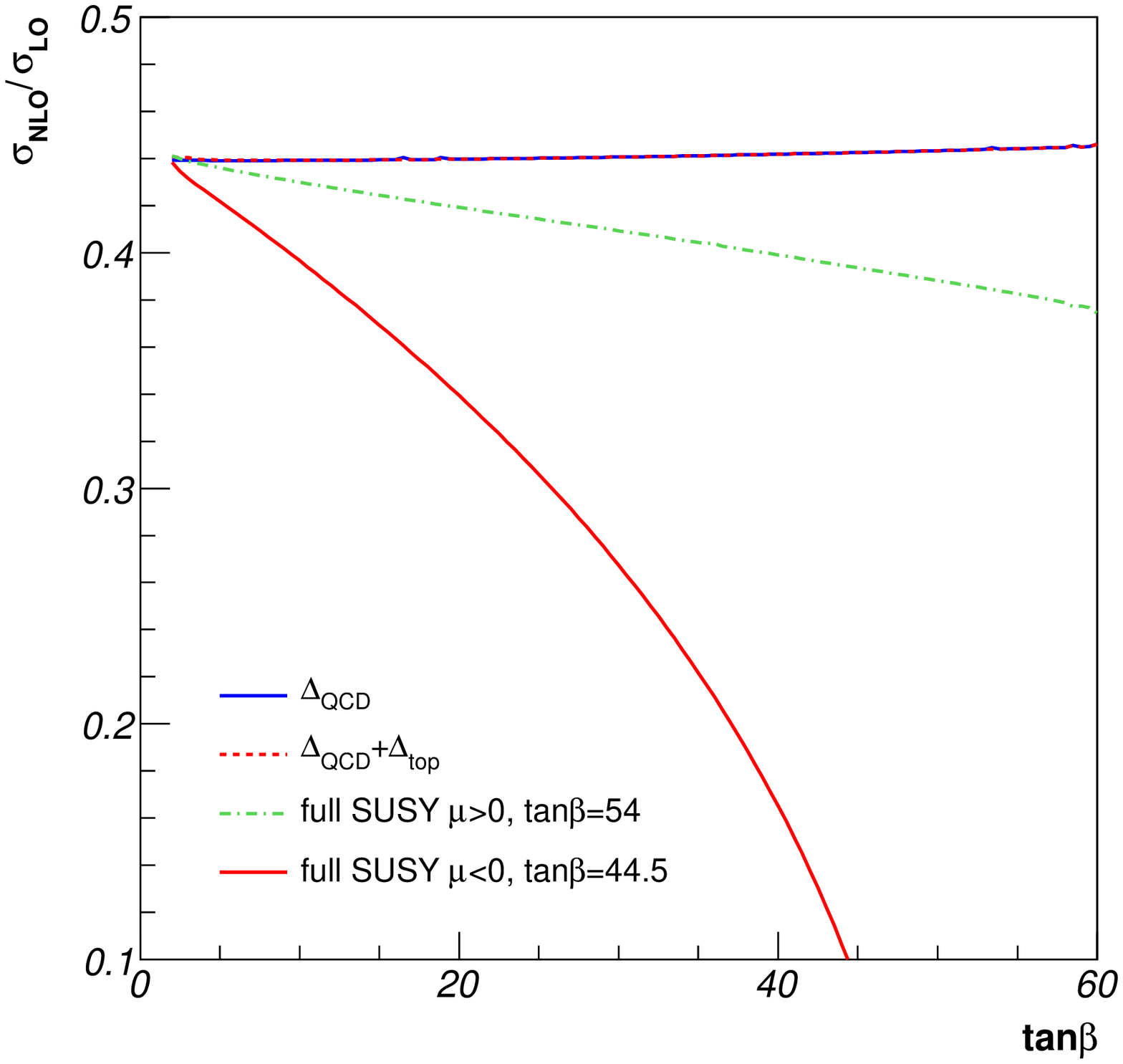}
	\hfill
	\includegraphics[scale=0.27]{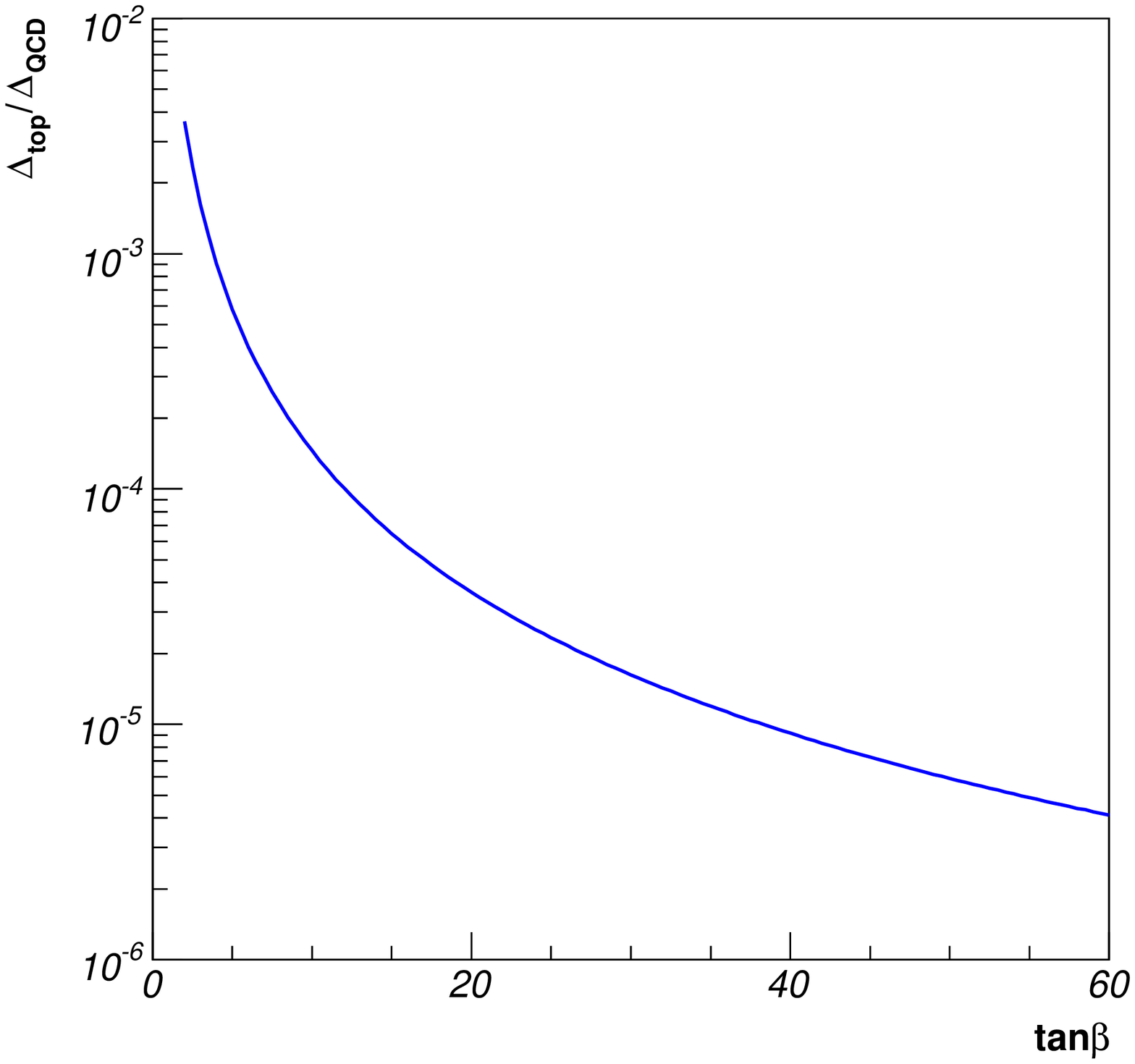}
	\hfill
	\includegraphics[scale=0.27]{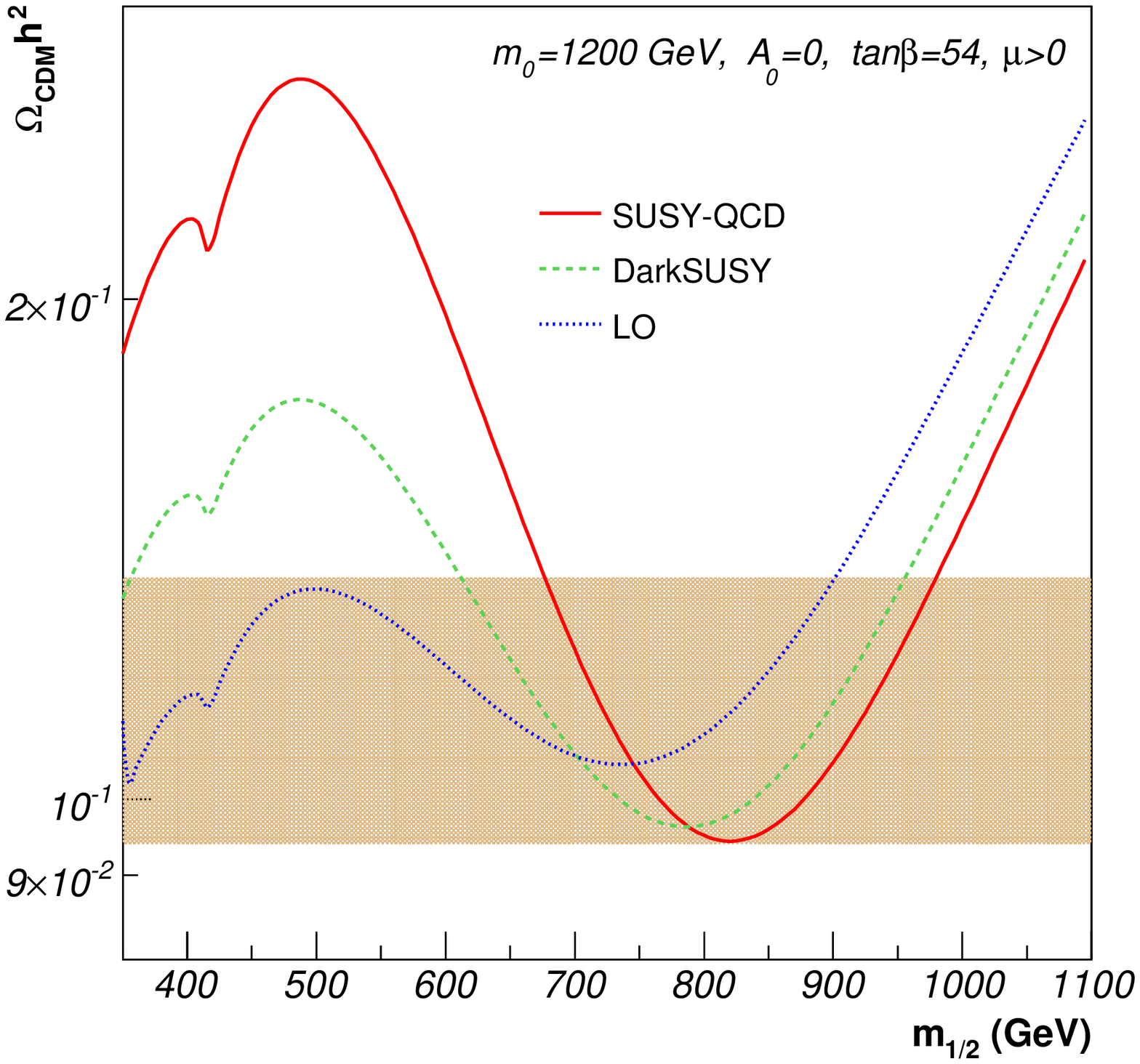}
\caption{Left and centre: NLO annihilation cross section in units of the LO 
	 cross section, including the QCD, QCD and top, and the full SUSY-QCD 
	 correction and ratio of top-quark loop correction and QCD correction 
	 as function of $\tan\beta$. Right: Dark matter relic density based on 
	 LO, NLO QCD, and full NLO SUSY-QCD calculation as function of $m_{1/2}$
	 for $m_0=1200$ GeV, $\mu<0$, $\tan\beta=44.5$ (slope indicated in the
	 left panel of Fig. \ref{fig4}).}
\label{fig3}
\end{figure*}

\begin{figure*}
	\includegraphics[scale=0.27]{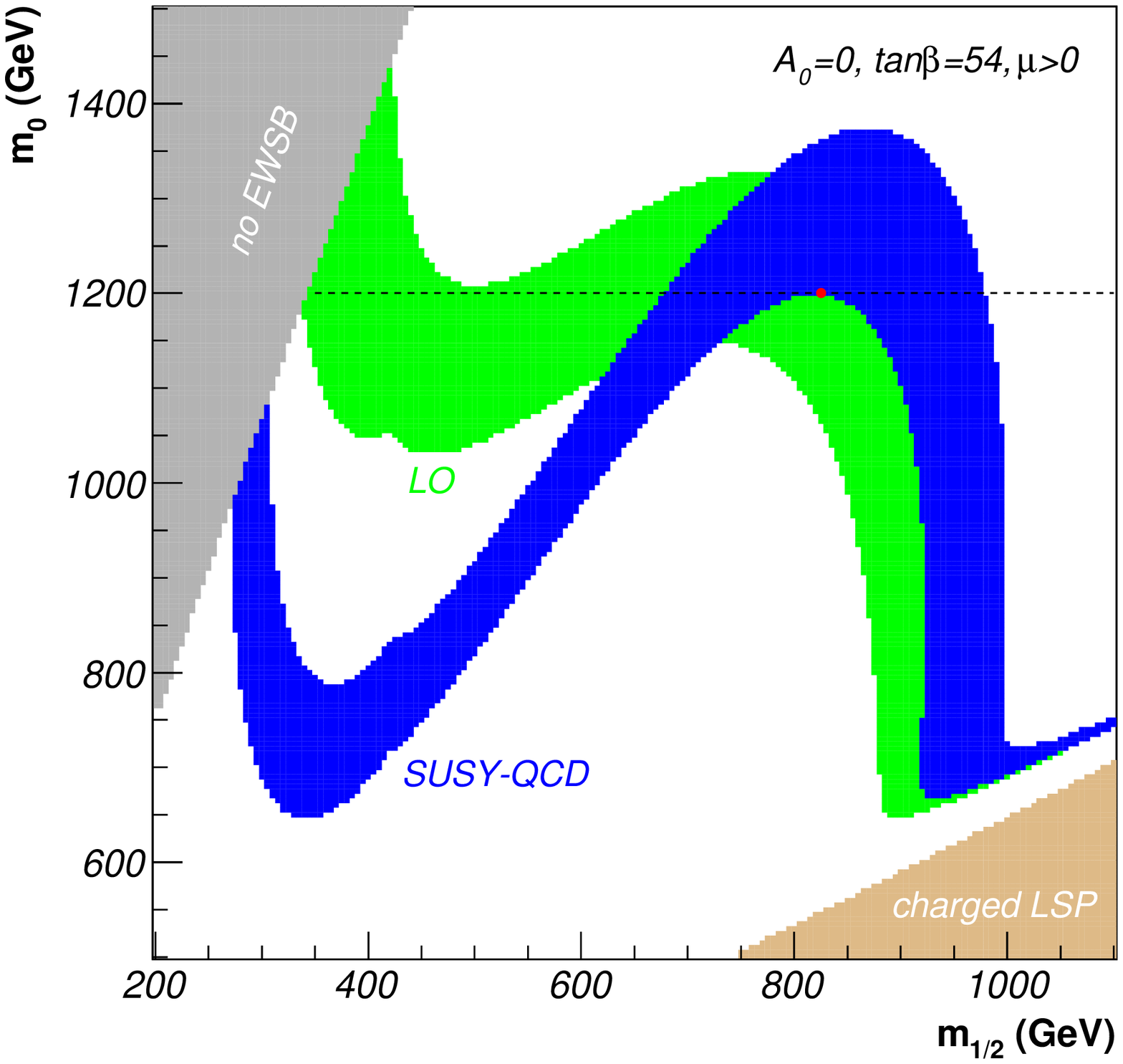}
	\hfill
	\includegraphics[scale=0.27]{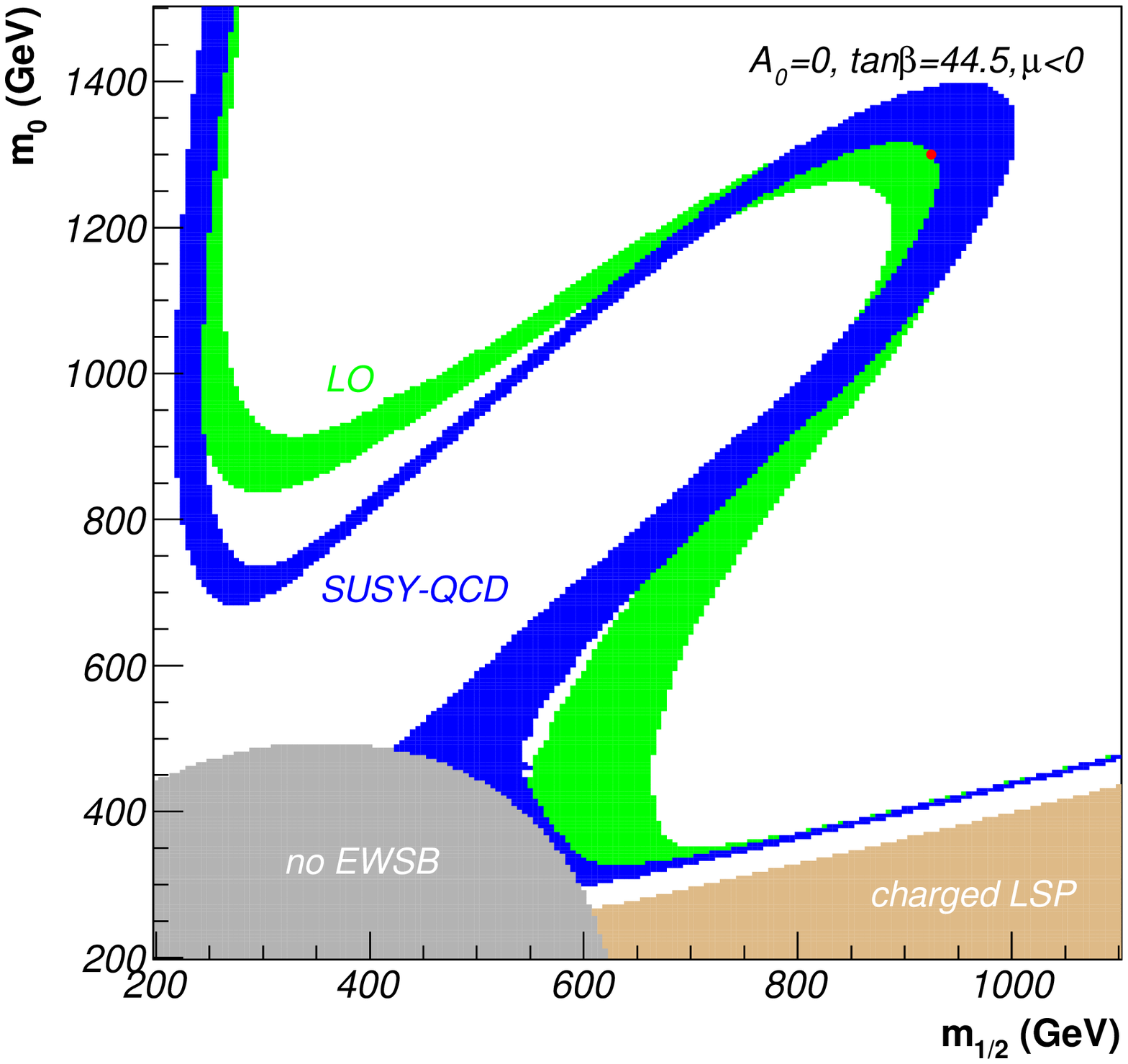}
	\hfill
	\includegraphics[scale=0.27]{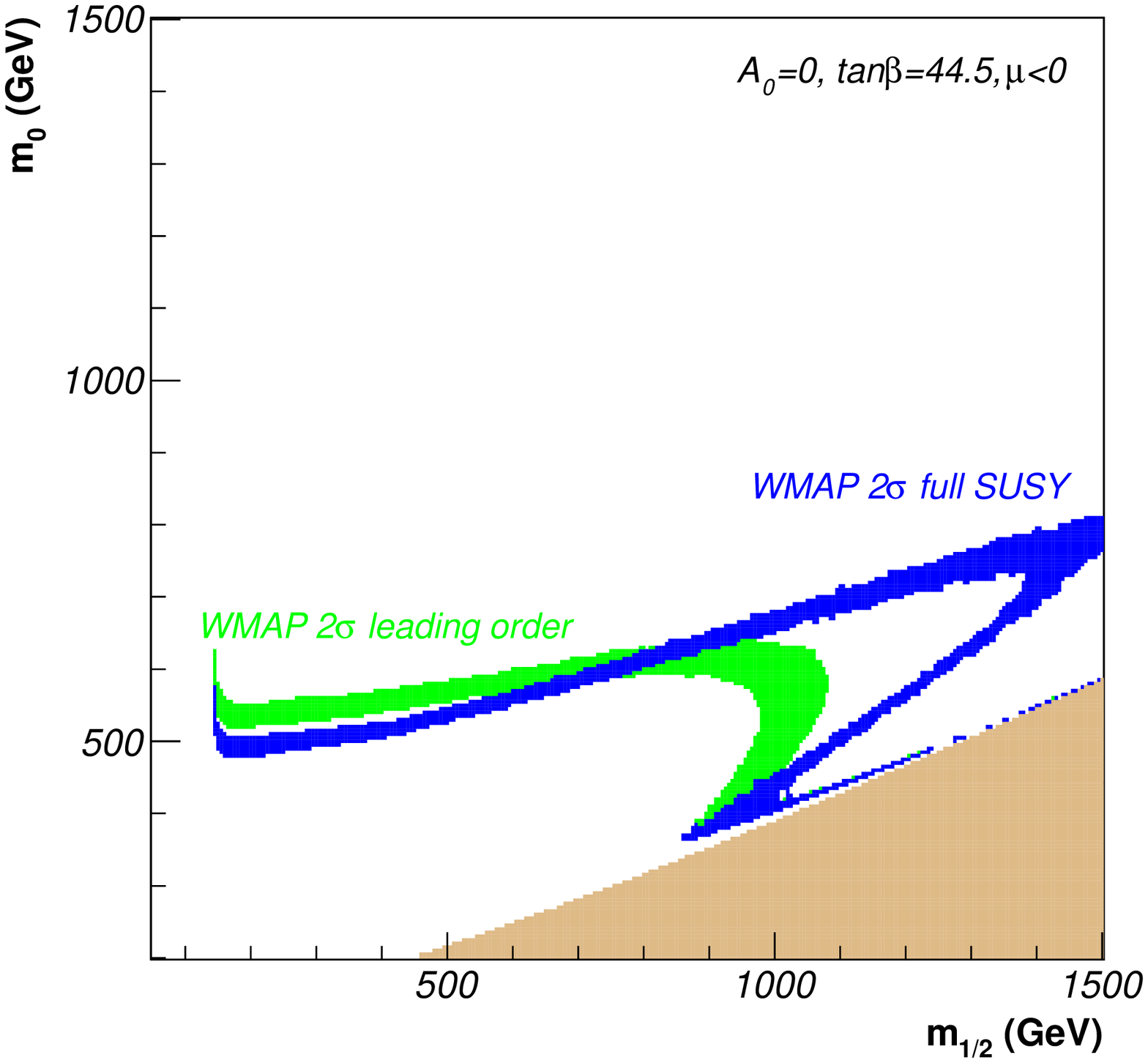}
\caption{Regions in the mSUGRA $m_0-m_{1/2}$ plane forbidden by a charged LSP / 
	 no EWSB and favoured by the observed cold dark matter relic density 
	 $\Omega_{\rm CDM}h^2$ including LO (light) and full SUSY-QCD (dark)
	 calculation. The corresponding SUSY spectrum has been obtained using 
	 {\tt SPheno} / {\tt FeynHiggs} for the left and centre panels, and 
	 using {\tt ISAJET} for the right panel.}
\label{fig4}
\end{figure*}

We first study the effect of the corrections on the annihilation cross section
relative to the process $\chi\chi\to A^0\to b\bar{b}$, which is presented in the
left-hand panel of Fig. \ref{fig3}. The corrected cross section $\sigma_{\rm
NLO}$ is shown normalized to the leading order (LO) cross section $\sigma_{\rm
LO}$, including either only the QCD diagrams, QCD and top-loop diagrams,
or the full SUSY-QCD correction. The QCD correction is independent of the SUSY
parameters,  while the top-quark loop correction is proportional to
$1/\tan^2\beta$ through  the $A^0tt$ coupling. The points, for which the
SUSY-QCD corrections are  evaluated, lie within the Higgs-funnel region and are
indicated by a point in the left and centre panels of Fig. \ref{fig4}, for
$\mu<0$ and $\mu>0$, respectively. It becomes clear that the annihilation cross
section is decreased by about a factor two by the QCD contribution, where the
principal contribution comes from the mass shift $m_b \to \bar{m}_b(s)$. The
top-quark loop contribution, as already mentioned, is negligible with respect to
the QCD contribution at large $\tan\beta$ such as in the {\it A-funnel} region.
In the centre panel of Fig. \ref{fig3} we see that $\Delta_{\rm top}$ accounts
for a few permille with respect to $\Delta_{\rm QCD}$ for $\tan\beta\simeq 2$,
and only for less than $10^{-5}$ for $\tan\beta > 40$. 

When adding the SUSY-QCD contribution $\Delta_{\rm SUSY}$, the annihilation
cross section receives another important correction, see left-hand panel of Fig.
\ref{fig3}. The SUSY correction decreases the cross section by another few
percent at very low $\tan\beta$ and up to another 40 (10) percent for $\mu<0$
($\mu>0$), which underlines the importance of the full correction in the
cosmological {\it A-funnel} region. The large difference between the correction
factors for $\mu>0$ and $\mu<0$ is mainly due to the mass renormalization, Eq.
(\ref{eq11}), that is directly affected by sgn($\mu$).

To evaluate the effect of the corrections on the cold dark matter relic density,
we have implemented our full SUSY-QCD correction, as described above, into the
{\tt DarkSUSY} package \cite{ref7}, which includes by default only the QCD 
corrections up to order ${\cal O}(\alpha_s^2)$ without their dependence on the 
energy scale $s$. Choosing either the pure Born or full SUSY-QCD calculation of 
the process $\chi\chi \to A^0 \to b\bar{b}$, we compare the resulting cold dark 
matter relic density $\Omega_{\rm CDM}h^2$ to the observational $2\sigma$ range 
in Eq. (\ref{eq1}) to determine the favoured regions in the $m_0-m_{1/2}$ plane
shown in the left-hand and centre panels of Fig. \ref{fig4}.
As expected, the corrections do not affect the cosmological focus point (very
low $m_{1/2}$), bulk (low $m_0$ and $m_{1/2}$) and coannihilation (low $m_0$)
regions, where the inspected process is suppressed with 
respect to other (co)annihilation channels. However, at large $\tan\beta=44.5$
(54) for $\mu<0$ ($\mu>0$), which still allow for electroweak symmetry breaking
(EWSB) in a large region of the scanned $m_0-m_{1/2}$ plane, the LO allowed
regions (light) are dramatically changed by the full SUSY corrections (dark),
which reduce $\sigma_{\rm eff}$ as discussed above. The increase in $\Omega_{\rm
CDM}$ is compensated by a shift towards smaller SUSY masses.

In the right-hand panel of Fig. \ref{fig3} we plot the relic density for
$m_0=1200$ GeV as function of $m_{1/2}$. The graph shows the contributions of
the different correction terms. The effect of of the ${\cal O}(\alpha_s^2)$ QCD
corrections already included in {\tt DarkSUSY} is considerably enhanced by our
newly included ${\cal O}(\alpha_s^3)$ QCD and ${\cal O}(\alpha_s)$ SUSY-QCD
corrections. Interesting effects due to the simultaneous correction of the Higgs
decay width $\Gamma_A$ are observed, e.g. the minimum of $\Omega_{\rm CDM}$ is
shifted towards the resonance $2m_{\chi}=m_A$. For a detailed discussion the
reader is referred to Ref. \cite{ref0}.

Finally we note a difference between the relic density calculated with our code
based on {\tt SPheno} and {\tt FeynHiggs} with respect to the one based on {\tt
ISAJET}, which is by default included in the {\tt DarkSUSY} package. The
right-hand panel of Fig. \ref{fig4} shows the {\it A-funnel} region for
$\mu<0$, which is to be found at smaller values of $m_0$. This difference should
be due to the fact that the spectrum calculation is especially sensible to the
spectrum generator for the neutralino and Higgs boson masses. In consequence,
the resonance condition $2m_{\chi}=m_A$, governing the {\it A-funnel} region,
can be shifted in an important way. However, the effect of the SUSY-QCD
corrections is the same, i.e. an important shift of the favoured
region with respect to the leading order result. For more details on the impact
of SUSY spectrum calculations on dark matter annihilation see Ref. \cite{ref23}.


\section{Conclusion}
\label{sec4}

We have presented a complete calculation of SUSY-QCD corrections to dark matter 
annihilation in the Higgs-funnel, resumming potentially large $\mu\tan\beta$ and 
$A_b$ contributions. We have demonstrated numerically that these corrections 
strongly influence the extraction of SUSY parameters from cosmological data 
and must therefore be included in common analysis tools like {\tt DarkSUSY} or 
{\tt micrOMEGAs}.



\begin{thebibliography}{999}

\bibitem{ref1}
D. Spergel {\it et al.} [WMAP collaboration], Astrophys. J. Suppl. \textbf{170} (2007) 377.

\bibitem{ref2}
R. Massey {\it et al.}, Nature \textbf{445} (2007) 286; \\
M. J. Lee {\it et al.}, Astrophys. J. \textbf{661} (2007) 278.

\bibitem{ref4}
A. Hamann {\it et al.}, Phys. Rev. D \textbf{75} (2007) 023522.

\bibitem{ref5}
G. Bertone {\it et al.}, Phys. Rept. \textbf{405} (2005) 279.

\bibitem{ref6}
G. Jungman {\it et al.}, Phys. Rept. \textbf{267} (1996) 195.

\bibitem{ref7}
P. Gondolo {\it et al.}, JCAP \textbf{0407} (2004) 008.

\bibitem{ref8}
G. B\'elanger {\it et al.}, Comput. Phys. Commun. \textbf{149} (2002) 103.

\bibitem{ref9}
M. S. Carena {\it et al.}, Nucl. Phys. B \textbf{429} (1994) 269.

\bibitem{ref10}
W. de Boer {\it et al.}, Astron. Astrophys. \textbf{444} (2005) 51; 
Phys. Lett. B \textbf{636} (2006) 13.

\bibitem{ref11}
L. Bergstrom {\it et al.}, JCAP \textbf{0605} (2006) 006.

\bibitem{ref0}
B. Herrmann and M. Klasen, arXiv:0709.0043 [hep-ph], submitted to PRL.

\bibitem{ref12}
M. Drees {\it et al.}, Phys. Rev. D \textbf{41} (1990) 1547; 
Phys. Lett. B \textbf{} (1990) 455; [Erratum-ibid. B \textbf{262} (1991) 497].

\bibitem{ref13}
E. Braaten {\it et al.}, Phys. Rev. D \textbf{22} (1980) 715.

\bibitem{ref14}
K. G. Chetyrkin, Phys. Lett.B \textbf{390} (1997) 309.

\bibitem{ref15}
K. G. Chetyrkin {\it et al.}, Nucl. Phys. B \textbf{461} (1996) 3.

\bibitem{ref16}
M. S. Carena {\it et al.}, Nucl. Phys. B \textbf{577} (2000) 88.

\bibitem{ref17}
J. Guasch {\it et al.}, Phys. Rev. D \textbf{68} (2003) 115001.

\bibitem{ref18}
A. Dabelstein, Nucl. Phys. B \textbf{456} (1995) 25; \\
J. A. Coarasa {\it et al.}, Phys. Lett. B \textbf{389} (1996) 312.

\bibitem{ref20}
W. Porod, Comput. Phys. Commun. \textbf{153} (2003) 275.

\bibitem{ref21}
S. Heinemeyer {\it et al.}, Comput. Phys. Commun. \textbf{124} (2000) 76.

\bibitem{ref22}
W. M. Yao {\it et al.} [Particle Data Group], J. Phys. G \textbf{33} (2006) 1.

\bibitem{ref23}
G. B\'elanger {\it et al.}, Phys. Rev. D \textbf{72} (2005) 015003.

\end{thebibliography}
\end{document}